\documentclass{sn-jnl}%

\usepackage{graphicx}
\usepackage{subcaption}
\usepackage{float}
\usepackage{rotating}
\usepackage{lscape}
\usepackage{capt-of}
\usepackage{multirow}
\usepackage{lscape}
\jyear{2021}%

\theoremstyle{thmstyleone}%
\theoremstyle{thmstyletwo}%

\theoremstyle{thmstylethree}%

\begin{document}

\title[A Dataset for Malware Classification]{A Multi-feature Dataset for Windows PE Malware Classification}


\author*[1]{\fnm{Muhammad Irfan} \sur{Yousuf}}\email{irfan.yousuf@uet.edu.pk}

\author[2]{\fnm{Izza} \sur{Anwer}}

\author[1]{\fnm{Tanzeela} \sur{Shakir}}
\author[1]{\fnm{Minahil} \sur{Siddiqui}}
\author[1]{\fnm{Maysoon} \sur{Shahid}}

\affil[1]{\orgdiv{Department of Computer Science (New Campus)}, \orgname{University of Engineering \& Technology}, \orgaddress{ \city{Lahore}, \country{Pakistan}}}

\affil[2]{\orgdiv{Department of Transportation Engineering and Management}, \orgname{University of Engineering \& Technology}, \orgaddress{\city{Lahore}, \country{Pakistan}}}


\abstract{This paper describes a multi-feature dataset for training machine learning classifiers for detecting malicious Windows Portable Executable (PE) files. The dataset includes four feature sets from 18,551 binary samples belonging to five malware families including Spyware, Ransomware, Downloader, Backdoor and Generic Malware. The feature sets include the list of DLLs and their functions, values of different fields of PE Header and Sections. First, we explain the data collection and creation phase and then we explain how did we label the samples in it using VirusTotal's services. Finally, we explore the dataset to describe how this dataset  can benefit the researchers for static malware analysis. The dataset is made public in the hope that it will help inspire machine learning research for malware detection.

}




\maketitle

\section{Introduction}

Malicious programs or malware have different types including Trojans, Spyware, Ransomware, Backdoor, and Downloaders to name a few \cite{Razak1}. Nowadays, computers have become an essential part of everyday life, and as a result, computer intruders are attacking
computers using different methods, or trying to use these computers as weapons. Malware analysis is the study of determining the bahavior of malware samples to prevent future attacks. In static malware analysis, we do not run the malware code in real rather we examine the file for malicious contents or signatures. Dynamic malware analysis executes the malicious program in a safe environment such as a sandbox and detects the run-time behavior of the program \cite{Ijaz1} 

Different malware have different purposes, e.g., obtaining personal or sensitive data such as bank accounts and passwords, gaining unauthorized access to a network, and blocking a few or all applications running on a system etc \cite{Shalaginov2018}. Generally, a malware attacks a specific platform or operating system, therefore, its behavior could depend on the underlying operating system being attacked \cite{Soliman1}. For example, a Windows Programmable Executable (PE) malware could access different Dynamic Link Libraries (DLLs) or call Application Programming Interface (API) routines to access files or other resources in a Windows environment. 

It has been observed that although malware have evolved structurally over the years, there are still some constant characteristics or features that help analysts to detect them. For example, if a malware is running on a Windows operating system, it will use some of the services of the operating system. The malware will access Windows DLLs to call different API functions. The set of DLLs accessed or the API functions called generate the malicious behavior. Similarly, the information of PE header and its different sections could help in detecting the malware. In short, if the malicious behavior is analyzed well, the detection of malware is possible. 

The main objective of this study is to collect a multi-feature dataset for Windows PE malware classification so that the researchers do not have to invest their time in collecting malware samples and extracting features from them. We focus on four features for static malware analysis including DLLs accessed, API functions called, PE Header and Sections information. We shared our dataset on GitHub site. We believe that the researchers working on cybersecurity or related fileds would benefit from this dataset. The dataset will fill the void in static malware analysis using machine learning. \cite{Aggarwal1,Balram1,Dama1,Wu1}.

\section{Data Collection}
\begin{figure*}[t]
	\centering
	\includegraphics[width = 120mm, height=110mm]{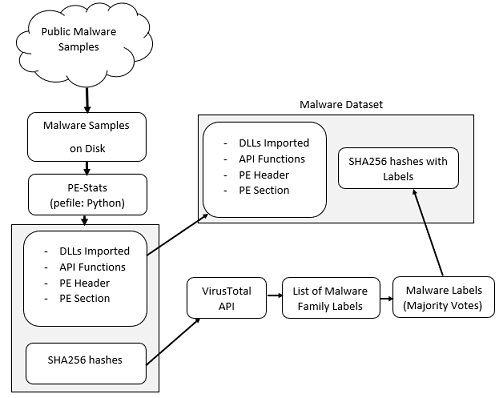}
	\caption{Data collection and labelling}
	\label{fig_DC}
\end{figure*}

The data were collected in two steps. Figure \ref{fig_DC} summarizes these steps. In the first step, we collected the data from MalwareBazaar Database \footnote{https://bazaar.abuse.ch/} using its API. Only Windows PE files were targeted in API calls and more than 20,000 samples were downloaded. We used pefile library \footnote{https://pypi.org/project/pefile/} of Python to extract PE statistics or features from those samples. The samples with incorrect or missing values in PE header were discarded. Similarly, we also discarded samples with code obfuscation \cite{Kane1}. After discarding unwanted samples, we have a total of 18,551 samples in our dataset. 

Ih the second step, we submitted the SHA256 hashes of all the samples to VirusTotal \footnote{https://www.virustotal.com/gui/home/upload} using its API for labeling the families of these samples. VirusTotal is a free service that allows us to analyze malware samples online. It has more than 70 antivirus engines or scanners for malware analyses. VirusTotal's public API can fetch the name of the malware family as classified by each antivirus. We pulled the family name of each sample as returned by all the antiviruses and saved them in a Comma Separated Values (CSV) file. 

\subsection{Dataset Creation}

The data set, as uploaded on GitHub \footnote{https://github.com/DA-Proj/PE-Malware-Dataset1} , has a very simple structure. We provide it as Comma Separated Values (CSV) files so that no specific software or tool is required to read it. We extracted four features including DLLs imported, APIs called, PE Header and Section information using pefile library. Each feature is presented in a separate CSV file. The first column of each CSV file contains SHA256 value of a sample whereas the second column contains its family name or label. We labeled the samples using majority voting method in which we label a sample with the family name that is returned by the majority of antivirus engines of VirusTotal. Based on the results of VirusTotal's public API, we classified the dataset into five classes as shown in Table ~\ref{Tab_Sample}. Each row of DLL and DLL Functions CSV files enlist the DLLs imported and API functions called by a sample respectively. Each row of Header CSV file provides values of 52 fields of header of each sample. Similarly, each row of Section CSV file gives values of ten fields of ten sections (a total of 100 features) of a sample.   

\begin{table}[t]
	\centering
	\caption{Distribution of malware according to their families}
	\begin{tabular}{p{2.5cm}p{1.2cm}p{7.5cm}}
		\hline
		\textbf{Malware Family} & \textbf{Count} & \textbf{Description} \\ \hline
		Generic Malware & 6,231 & Generic malware that do not fit in any class below. \\ 
		Spyware & 5,766 & It monitors activities on an infected system and steals sensitive information such as passwords and bank details. \\ 
		Downloader & 2,438 & It helps in downloading other malwares onto an infected system. \\ 
		Ransomware & 2,376 & It prevents users from accessing their system or personal files and demands ransom payment in order to regain access. \\ 
		Backdoor & 1,740 & A trojan that negates normal authentication procedures to access a system. It grants remote access to the resources under attcak. \\ \hline
		\textbf{Total} & \textbf{18,551} &  \\ \hline
	\end{tabular}
	\label{Tab_Sample}
\end{table}

\section{Dataset Description}
The dataset contains four features extracted from 18,551 malware samples. We detail the four features as follows.

\subsection{Features}
The dataset contains the following four features of Windows PE malware samples. These are static malware analysis features, i.e., we did not run the samples in a sandbox. All these features have been extracted using Python's pefile library.  \\
\textbf{List of Imported DLLs: } We extracted the list of all the DLLs imported by a sample. Malware writers use Windows DLLs to avoid writing the code that they think is available in a Windows DLL. A program can be characterized by the set of DLLs it imports. Therefore, this feature can be used for classification of malware.\\
\textbf{List of API calls: } This feature contains the list of all the API functions called by a malware. This feature supplements the first feature and it can be used to improve the classification accuracy.\\
\textbf{PE Header: } PE header contains useful data about the sample under study. A PE file contains a number of headers such as COFF file header, MS-DOS Stub, and optional header etc. We collected the values of 52 fields of PE Header as our third feature set. The list of these 52 fields is given in Table \ref{Tab_Header}.\\
\textbf{PE Section: } Many sections such as code section (.text), data section (.data, .rdata), and resource section (.rsrc) etc. are part of Windows PE files. We extracted the values of ten fields of each of ten sections (a total of 100 section values) as our fourth feature set. The details of these sections and values are given in Table ~\ref{Tab_section}.

\begin{table}[t]
	\centering
	\caption{PE Header fields in the 3rd feature set}
	\begin{tabular}{|l|l|}
		\hline
		\textbf{DOS Header} & \textbf{Optional Header} \\ \hline
		e\_magic & Magic \\ \hline
		e\_cblp & MajorLinkerVersion \\ \hline
		e\_cp & MinorLinkerVersion \\ \hline
		e\_crlc & SizeOfCode \\ \hline
		e\_cparhdr & SizeOfInitializedData \\ \hline
		e\_minalloc & SizeOfUninitializedData \\ \hline
		e\_maxalloc & AddressOfEntryPoint \\ \hline
		e\_ss & BaseOfCode \\ \hline
		e\_sp & ImageBase \\ \hline
		e\_csum & SectionAlignment \\ \hline
		e\_ip & FileAlignment \\ \hline
		e\_cs & MajorOperatingSystemVersion \\ \hline
		e\_lfarlc & MinorOperatingSystemVersion \\ \hline
		e\_ovno & MajorImageVersion \\ \hline
		e\_oemid & MinorImageVersion \\ \hline
		e\_oeminfo & MajorSubsystemVersion \\ \hline
		e\_lfanew & MinorSubsystemVersion \\ \hline
		- & Reserved1 \\ \hline
		- & SizeOfImage \\ \hline
		- & SizeOfHeaders \\ \hline
		\textbf{File Header} & CheckSum \\ \hline
		Machine & Subsystem \\ \hline
		NumberOfSections & DllCharacteristics \\ \hline
		TimeDateStamp & SizeOfStackReserve \\ \hline
		PointerToSymbolTable & SizeOfHeapReserve \\ \hline
		NumberOfSymbols & SizeOfHeapCommit \\ \hline
		SizeOfOptionalHeader & LoaderFlags \\ \hline
		Characteristics & NumberOfRvaAndSizes \\ \hline
	\end{tabular}
	\label{Tab_Header}
\end{table}

\begin{table}[htb]
	\centering
	\caption{PE sections and fields of 4th feature set}
	\begin{tabular}{p{3cm}p{8cm}}
		\hline
		\textbf{Section Name} & \textbf{Description} \\ \hline
		.text & This section contains the executable code.   It also contains program entry point. \\ 
		.data & This section contains initialized data of a program. \\ 
		.rdata & It contains data that is to be only readable, such as literal strings, and constants. \\ 
		.bss & It represents uninitialized data to reduce the size of executable file. \\ 
		.idata & The .idata section contains  information about imported functions. \\ 
		.edata & This section contains information about symbols that other images can access through dynamic linking. \\ 
		.rsrc & This  resource-container section contains resource information. \\ 
		.reloc & Relocation information is saved in this section. \\ 
		.tls & TLS stands for Thread Local Storage. Each thread running in Windows uses its own storage called TLS. \\ 
		.pdata & The .pdata section contains an array of function table entries that are used for exception handling. \\ \hline
		\textbf{Field Name} & \textbf{Description} \\ \hline
		Name & An 8-byte encoded string contains name of the section. \\ 
		Misc\_VirtualSize & The total size of the section when loaded into memory. \\ 
		VirtualAddress & The address of the first byte of a section. \\ 
		SizeOfRawData & The size of the section. \\ 
		PointerToRawData & The file pointer to the first page of the section within the COFF file. \\ 
		PointerToRelocations & The file pointer to the beginning of relocation entries for the section. \\ 
		PointerToLinenumbers & The file pointer to the beginning of line-number entries for the section. \\ 
		NumberOfRelocations & The number of relocation entries for the section. \\ 
		NumberOfLinenumbers & The number of line-number entries for the section.  \\ 
		Characteristics & The flags that describe the characteristics of the section. \\ \hline
	\end{tabular}
	\label{Tab_section}
\end{table}

\subsection{Exploring Dataset}
Table ~\ref{Tab_DLL} enlist the top-30 DLLs imported by each malware family. We see that top-5 DLLs are same for all families, however, as we move down the list each malware family starts importing task specific DLLs. The top-5 DLLs provide basic functionalities such as memory, files and hardware access and almost all programs whether malware or benign need these basic functionalities. Then, we see that Spyware and Ransomware import crypt32.dll for Certificate and Cryptographic Messaging functions whereas Downloader malware family does not import this DLL. It means that imported DLLs can be used as a feature for classification tasks as each malware family will probably import a different set of DLLs. Similarly,  Table ~\ref{Tab_DLL_Func} enlist top-30 API functions called by each malware family. We see that each malware family calls a different set of API functions to accomplish its specific task and makes the list of API function calls an important feature for classification. We believe that combining these two features can enhance the accuracy of a classification model. 

To summarize, a total of 13,835 unique API functions from 427 unique DLLs are called by the malware samples in this dataset. There are 52 features in the Header feature set whereas the Sections feature set contains 100 features. Overall, the dataset contains 14,414 features of Windows PE files.

\begin{landscape}
	\begin{table}[ht]
		\centering
		\caption{Top 30 DLLs imported by each malware family.}
		\begin{tabular}{|l|l|l|l|l|}
			\hline
			\textbf{Generic Malware} & \textbf{Backdoor} & \textbf{Downloader} & \textbf{Ransomware} & \textbf{Spyware} \\ \hline
			mscoree.dll & mscoree.dll & kernel32.dll & kernel32.dll & kernel32.dll \\ \hline
			kernel32.dll & kernel32.dll & mscoree.dll & user32.dll & user32.dll \\ \hline
			user32.dll & user32.dll & user32.dll & advapi32.dll & mscoree.dll \\ \hline
			advapi32.dll & advapi32.dll & advapi32.dll & ole32.dll & ole32.dll \\ \hline
			ole32.dll & ole32.dll & ole32.dll & mscoree.dll & advapi32.dll \\ \hline
			gdi32.dll & oleaut32.dll & gdi32.dll & gdi32.dll & gdi32.dll \\ \hline
			shell32.dll & gdi32.dll & shell32.dll & shell32.dll & oleaut32.dll \\ \hline
			oleaut32.dll & shell32.dll & comctl32.dll & oleaut32.dll & shell32.dll \\ \hline
			comctl32.dll & comctl32.dll & oleaut32.dll & msvcrt.dll & comctl32.dll \\ \hline
			msvcrt.dll & msvcrt.dll & msvcrt.dll & comctl32.dll & msvcrt.dll \\ \hline
			ws2\_32.dll & version.dll & ws2\_32.dll & winhttp.dll & winhttp.dll \\ \hline
			version.dll & ws2\_32.dll & version.dll & version.dll & version.dll \\ \hline
			shlwapi.dll & winmm.dll & comdlg32.dll & ws2\_32.dll & ws2\_32.dll \\ \hline
			comdlg32.dll & comdlg32.dll & winmm.dll & shlwapi.dll & comdlg32.dll \\ \hline
			winhttp.dll & wininet.dll & shlwapi.dll & ntdll.dll & shlwapi.dll \\ \hline
			winmm.dll & shlwapi.dll & wininet.dll & wininet.dll & odbc32.dll \\ \hline
			wininet.dll & winspool.dll & api-ms-win-crt-runtime-l1-1-0.dll & comdlg32.dll & winmm.dll \\ \hline
			winspool.dll & winhttp.dll & api-ms-win-crt-stdio-l1-1-0.dll & winmm.dll & gdiplus.dll \\ \hline
			msvbvm60.dll & netapi32.dll & api-ms-win-crt-locale-l1-1-0.dll & gdiplus.dll & dwrite.dll \\ \hline
			msvcr90.dll & gdiplus.dll & vcruntime140.dll & winspool.dll & wininet.dll \\ \hline
			gdiplus.dll & ntdll.dll & api-ms-win-crt-heap-l1-1-0.dll & netapi32.dll & winspool.dll \\ \hline
			crypt32.dll & psapi.dll & api-ms-win-crt-math-l1-1-0.dll & wsock32.dll & netapi32.dll \\ \hline
			netapi32.dll & msimg32.dll & api-ms-win-crt-convert-l1-1-0.dll & iphlpapi.dll & ntdll.dll \\ \hline
			ntdll.dll & iphlpapi.dll & msvcp140.dll & crypt32.dll & ole32.dll \\ \hline
			wsock32.dll & oleacc.dll & api-ms-win-crt-utility-l1-1-0.dll & mpr.dll & msvbvm60.dll \\ \hline
			cgraph.dll & ntoskrnl.dll & api-ms-win-crt-string-l1-1-0.dll & mfc42.dll & rpcrt4.dll \\ \hline
			api-ms-win-crt-runtime-l1-1-0.dll & crypt32.dll & api-ms-win-crt-time-l1-1-0.dll & psapi.dll & crypt32.dll \\ \hline
			api-ms-win-crt-heap-l1-1-0.dll & crtdll.dll & api-ms-win-crt-filesystem-l1-1-0.dll & api-ms-win-crt-runtime-l1-1-0.dll & wsock32.dll \\ \hline
			vcruntime140.dll & mpr.dll & winspool.dll & vcruntime140.dll & urlmon.dll \\ \hline
			api-ms-win-crt-stdio-l1-1-0.dll & wsock32.dll & api-ms-win-crt-environment-l1-1-0.dll & api-ms-win-crt-heap-l1-1-0.dll & msimg32.dll \\ \hline
		\end{tabular}
		\label{Tab_DLL}
	\end{table}
\end{landscape}

\begin{landscape}
	\begin{table}[htb]
		\centering
		\caption{Top 30 API functions called by each malware family.}
		\scalebox{0.97}{
		\begin{tabular}{|l|l|l|l|l|l|}
			
			\hline
			\textbf{Generic Malware} & \textbf{Backdoor} & \textbf{Downloader} & \textbf{Ransomware} & \textbf{Spyware} \\ \hline
			\_corexemain & \_corexemain & \_corexemain & getprocaddress & getprocaddress \\ \hline
			getprocaddress & getprocaddress & getprocaddress & getlasterror & getlasterror \\ \hline
			getlasterror & getlasterror & getlasterror & exitprocess & exitprocess \\ \hline
			getcurrentprocess & exitprocess & writefile & getcurrentprocess & getcurrentthreadid \\ \hline
			exitprocess & writefile & closehandle & writefile & closehandle \\ \hline
			closehandle & closehandle & exitprocess & closehandle & writefile \\ \hline
			writefile & sleep & getcurrentprocess & getcurrentthreadid & multibytetowidechar \\ \hline
			multibytetowidechar & getcurrentthreadid & sleep & multibytetowidechar & getcurrentprocess \\ \hline
			sleep & getcurrentprocess & multibytetowidechar & sleep & entercriticalsection \\ \hline
			getcurrentthreadid & multibytetowidechar & widechartomultibyte & widechartomultibyte & leavecriticalsection \\ \hline
			widechartomultibyte & entercriticalsection & freelibrary & entercriticalsection & widechartomultibyte \\ \hline
			entercriticalsection & leavecriticalsection & getmodulehandlea & leavecriticalsection & deletecriticalsection \\ \hline
			leavecriticalsection & deletecriticalsection & getmodulehandlew & deletecriticalsection & \_corexemain \\ \hline
			deletecriticalsection & widechartomultibyte & readfile & unhandledexceptionfilter & getstdhandle \\ \hline
			unhandledexceptionfilter & unhandledexceptionfilter & gettickcount & terminateprocess & getcurrentprocessid \\ \hline
			terminateprocess & terminateprocess & getcurrentthreadid & getcurrentprocessid & unhandledexceptionfilter \\ \hline
			getcurrentprocessid & getcurrentprocessid & getmodulefilenamew & getstdhandle & sleep \\ \hline
			setunhandledexceptionfilter & getstdhandle & createfilew & setunhandledexceptionfilter & terminateprocess \\ \hline
			getstdhandle & getmodulehandlea & entercriticalsection & queryperformancecounter & setunhandledexceptionfilter \\ \hline
			queryperformancecounter & queryperformancecounter & leavecriticalsection & getsystemtimeasfiletime & setlasterror \\ \hline
			getsystemtimeasfiletime & loadlibrarya & deletecriticalsection & getmodulehandlew & queryperformancecounter \\ \hline
			gettickcount & freelibrary & getstdhandle & freelibrary & getsystemtimeasfiletime \\ \hline
			getmodulehandlew & setunhandledexceptionfilter & unhandledexceptionfilter & setlasterror & getacp \\ \hline
			freelibrary & gettickcount & regclosekey & getmodulefilenamew & getcpinfo \\ \hline
			setlasterror & getsystemtimeasfiletime & getcurrentprocessid & getacp & tlsgetvalue \\ \hline
			getcommandlinea & getmodulefilenamea & setfilepointer & tlsgetvalue & heapfree \\ \hline
			getacp & getcommandlinea & findclose & getcpinfo & heapalloc \\ \hline
			getmodulehandlea & tlsgetvalue & terminateprocess & heapalloc & getfiletype \\ \hline
			heapalloc & getmodulehandlew & setunhandledexceptionfilter & heapfree & isdebuggerpresent \\ \hline
			heapfree & readfile & queryperformancecounter & getfiletype & raiseexception \\ \hline
		\end{tabular}}
		\label{Tab_DLL_Func}
	\end{table}
	
\end{landscape}

\section*{Acknowledgment}
This project was funded by KIST School Partnership Project, an initiative by Korea Institute of Science and Technology (KIST), Seoul, Republic of Korea, to support its alumni.

\bibliographystyle{sn-mathphys}      
\bibliography{MalwareDataset_bib}   



\end{document}